\documentclass[preprint,12pt]{elsarticle}
\usepackage{amsmath}
\usepackage{amsfonts}
\usepackage{titlesec}
\usepackage[symbol]{footmisc}
\usepackage{arydshln}
\usepackage{hyperref}
\titleclass{\subsubsubsection}{straight}[\subsection]
\newcounter{subsubsubsection}[subsubsection]
\renewcommand\thesubsubsubsection{\thesubsubsection.\arabic{subsubsubsection}}

\titleformat{\subsubsubsection}
  {\normalfont\normalsize}{\thesubsubsubsection}{1em}{}
\titlespacing*{\subsubsubsection}
{0pt}{3.25ex plus 1ex minus .2ex}{1.5ex plus .2ex}

\makeatletter
\renewcommand\paragraph{\@startsection{paragraph}{5}{\z@}%
  {3.25ex \@plus1ex \@minus.2ex}%
  {-1em}%
  {\normalfont\normalsize}}
\renewcommand\subparagraph{\@startsection{subparagraph}{6}{\parindent}%
  {3.25ex \@plus1ex \@minus .2ex}%
  {-1em}%
  {\normalfont\normalsize}}
\def\toclevel@subsubsubsection{4}
\def\toclevel@paragraph{5}
\def\toclevel@subparagraph{6}
\def\l@subsubsubsection{\@dottedtocline{4}{7em}{4em}}
\def\l@paragraph{\@dottedtocline{5}{10em}{5em}}
\def\l@subparagraph{\@dottedtocline{6}{14em}{6em}}
\makeatother

\usepackage{amssymb}
\usepackage[T1]{fontenc}
\usepackage{graphicx}
\usepackage{natbib}         
\usepackage{hyperref}       

\usepackage{longtable}
\usepackage{array}
\usepackage{booktabs} 
\usepackage[margin=1in]{geometry}

\date{October 15, 2024}

\begin{document}

\begin{frontmatter}

\setcounter{footnote}{0}
\title{Interpretable Machine Learning Model for Predicting Activist Investment Targets.}


\author[1]{Minwu Kim\thanks{Email: \texttt{mwk300@nyu.edu} \newline ORCID: \href{ https://orcid.org/0009-0005-8104-2142}{0009-0005-8104-2142}}}
\author[1]{Sidahmed Benabderrahmane\thanks{Email: \texttt{sidahmed.benabderrahmane@nyu.edu}}}
\author[1]{Talal Rahwan\thanks{Email: \texttt{talal.rahwan@nyu.edu} \newline ORCID: \href{https://orcid.org/0000-0003-0070-0667}{0000-0003-0070-0667}}}

\affiliation[1]{organization={New York University Abu Dhabi},            addressline={PO Box 129188, Saadiyat Island},city={Abu Dhabi}, country={United Arab Emirates}}

\begin{abstract}
This research presents a predictive model to identify potential targets of activist investment funds—entities that acquire significant corporate stakes to influence strategic and operational decisions, ultimately enhancing shareholder value. Predicting such targets is crucial for companies aiming to mitigate intervention risks, activist funds seeking optimal investments, and investors looking to leverage potential stock price gains. Using data from the Russell 3000 index from 2016 to 2022, we evaluated 123 model configurations incorporating diverse imputation, oversampling, and machine learning techniques. Our best model achieved an AUC-ROC of 0.782, demonstrating its capability to effectively predict activist fund targets. To enhance interpretability, we employed the Shapley value method to identify key factors influencing a company’s likelihood of being targeted, highlighting the dynamic mechanisms underlying activist fund target selection. These insights offer a powerful tool for proactive corporate governance and informed investment strategies, advancing understanding of the mechanisms driving activist investment decisions.

\end{abstract}

\begin{keyword}
Activist Funds \sep Shareholder Activism \sep Corporate Finance \sep Machine Learning \sep Explainable AI \sep SHAP
\end{keyword}

\end{frontmatter}


\section{Introduction}
Activist funds are a distinct class of investment funds that seek to effect change within their portfolio companies, not just through investment but by actively influencing management and strategic decisions. These funds differentiate themselves from traditional investment vehicles by adopting a hands-on approach, leveraging their stakes in companies to advocate for changes in corporate governance, strategy, management, and operations \citep{Aluchna2023}. Their ultimate goal is to unlock shareholder value that might not be realized under the current management's approach.

The participation of activists has been growing in recent years, particularly after the financial scandals and the financial crisis of 2007–2008 \citep{DesJardineDurand2020}. In 2018, nearly one campaign launched against a new target every day \citep{Lazard2018}. Thanks to regulatory reforms that have made shareholder engagement more accessible and weakened corporate defenses, activists can now more easily challenge management to enhance shareholder benefits \citep{Briggs2007}. The motivations behind activist investing are diverse. Romito identifies four main types of activist investors \citep{Romito2015}. Some focus on enhancing returns by advising management to issue debt or redirect free cash flow toward dividends or stock buybacks. Others advocate for merging similar competitors to capitalize on economies of scale. Another group prefers to enhance profitability by separating mismatched divisions within a company. Additionally, some investors aim to streamline operations by maintaining only the most profitable units and spinning off less effective ones. Beyond these financial goals, certain socially responsible funds also promote environmental, social, and governance (ESG) improvements, aiming to foster ethical business practices \citep{Barko2022}.

Studies have demonstrated the benefits of activist interventions in terms of market value, profitability, and governance, although there is ongoing debate about how these benefits vary over time \citep{DesJardineDurand2020}, \citep{Flammer2015}, \citep{Bebchuk2015}. Bebchuk et al. conducted an analysis of approximately 2,000 activist interventions from 1994 to 2007, revealing that activist funds often target underperforming companies with stock returns significantly lower than those of industry peers. However, in the five years following these interventions, the performance of the targeted companies generally improved, with them closing two-thirds of their performance gap in terms of return on assets compared to their peers \citep{Bebchuk2015}. Conversely, Desjardine and Durand analyzed 1,324 campaigns between 2000 and 2016 and found that while shareholder benefits were initially apparent and significant, they were short-lived. These gains often came at a long-term cost to other stakeholders, manifested in reduced cash flow and lower investment spending \citep{DesJardineDurand2020}.

Recognizing potential targets of these activist funds is therefore crucial for a wide array of stakeholders in today's market regime. It enables corporations to prepare and defend against unwanted interventions, guides activist investors in selecting firms where they can significantly impact, and offers retail investors opportunities for profit through potential stock price appreciations following activist campaigns.

Our research focuses on creating a model that can accurately predict which companies will become targets of activist funds. We analyze data from firms listed on the Russell 3000 index from 2016 to 2022, along with information on activist campaigns. Our analysis
includes a wide range of variables encompassing traditional measures of valuation and operations, as well as alternative data points related to ownership structures, governance, and technical indicators. By experimenting with 123 distinct combinations of imputation,
oversampling, and machine learning techniques, we determine the most effective model. Furthermore, we utilize Shapley values \citep{LundbergLee2017} to pinpoint the key factors that influence a company's likelihood of being targeted by an activist fund. This
approach not only enhances the predictive accuracy of our model but also provides deeper insights into the dynamics of shareholder activism.

\section{Literature Review}
Activist funds have long been known to focus on traditional valuation metrics when selecting targets, as revealed by various studies. Pfirrmann and Eichner's study based on 30 campaigns between 2013 and 2019, reveals a preference for relative valuation over intrinsic valuations, primarily utilizing Enterprise Value/EBITDA (EV/EBITDA) and Price/Earnings (P/E) ratios \citep{PfirrmannEichner2023}. Consistent with findings by Brav et al. \citep{BravJiangPartnoyThomas2008} and Boyson and Mooradian \citep{BoysonMooradian2011}, it is observed that firms targeted by activists often exhibit lower valuations compared to their fundamental values. Operationally, these firms show robust cash flows from operations and sales growth rates. Furthermore, their asset returns outperform those of their competitors, despite their stock performance trailing behind the broader market. The analysis also highlights that activist investors typically secure a significant stake in these firms, purchasing between 5\% to 10\% of equity shares on the market once they have identified their targets.

In order to reconfirm the preference for traditional valuation metrics, the evaluation methods used by institutional investors, not just activist funds, were also examined, and similar trends were confirmed in various studies. Bancel and Mittoo conducted a survey among 365 European finance experts holding CFAs or similar qualifications, finding that the most favored valuation techniques are Discounted Cash Flows (DCF) and relative valuation methods, particularly emphasizing the EV/EBITDA and P/E ratios as the top choices for multiples \citep{BancelMittoo2014}. Similarly, Mukhlynina and Nyborg's (2016) study involving valuation specialists, such as investment bankers, consultants, and private equity professionals, revealed that a significant portion of these experts employ both multiples and DCF in their valuation practices, with 47\% admitting to using both methods but predominantly favoring multiples \citep{MukhlyninaNyborg2016}.

Several studies have explored the evolving focus of activist funds on environmental, social, and governance (ESG) issues. A study by Zhu revealed that activist interventions not only improve the valuations and operations of a company but also steer management practices in a more shareholder-friendly direction. These interventions have led to increased shareholder returns through higher buyback activities and increased dividend payout ratios, as well as reductions in CEO compensation \citep{Zhu2020}. In addtion, Barko et al. demonstrated that socially responsible activist funds often choose large, high-profile companies with good financial performance and liquidity but poor ESG scores. Their research indicates that activism focused on corporate social responsibility typically enhances both ESG practices and corporate sales performance, suggesting that ethical investing can coincide with robust financial returns \citep{Barko2022}. The trend towards social responsibility in activism is further supported by research from Albuquerque et al., who developed a model demonstrating that firms decide to engage in CSR activities. Their study noted that over 3,000 institutions, managing assets upwards of 90 trillion USD, have committed to integrating CSR into their due diligence processes \citep{Albuquerque2019}. Additionally, research by Francis et al. underscored the significance of governance-related criteria in target selection, revealing that activist funds are approximately 52\% more likely to choose firms with female CEOs. These targets led by women are more cooperative towards activists and tend to see greater improvements in market and operational performance during campaigns \citep{Francis2021}.

Our research builds upon previous studies to provide fresh insights into shareholder activism. Specifically, the research have 4 main contributions:
\begin{itemize}
    \item First, we explore a wide range of company characteristics to assess their likelihood of being targeted by activist funds. While traditional variables like valuations and operations have been extensively studied, we incorporate less-explored factors such as governance, ownership, technical, and social aspects. This offers a more comprehensive view of the companies that attract activist interest.
    \item Second, we employ advanced machine learning techniques to develop a predictive model. Unlike traditional econometric methods like logistic regression, which are limited in capturing the complexity of activist decision-making, machine learning models can exploit non-linear and non-parametric patterns, resulting in significantly improved prediction accuracy.
    \item Third, we go beyond conventional methods of analyzing variable importance, like standard coefficient analysis, by using SHAP (SHapley Additive exPlanations). This game-theory-based approach allows us to measure the contribution of different variables at the individual instance level, providing deeper insights into how activists select their targets.
    \item 
    Fourth, we analyze the most recent data from the US equity market, covering the period from 2016 to 2022. Given the dynamic nature of financial markets and frequent regime changes, focusing on this latest data enables us to capture the most relevant and up-to-date trends in activist fund behavior.
\end{itemize}

\section{Methodology}
\begin{figure}[h]
\includegraphics[width=1\textwidth]{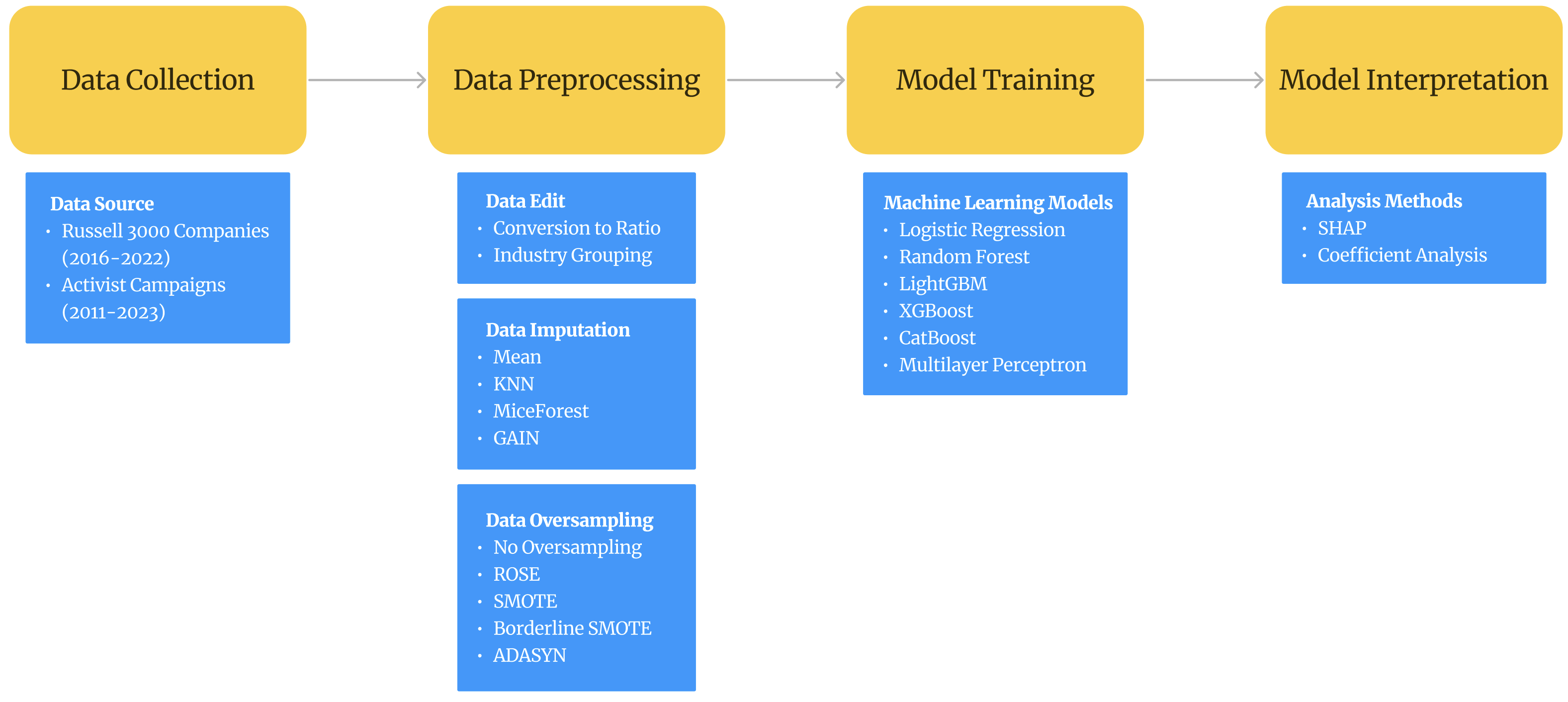}
\caption{Overview of the methodology.} \label{pipeline}
\end{figure}

Fig. \ref{pipeline} provides an overview of the proposed pipeline for activist investment target prediction, consisting of several stages. It begins by collecting data from Russell 3000 companies (2016-2022) and activist campaigns (2011-2023), featuring diverse variables. In preprocessing, non-ratio variables are transformed into ratios and selected variables are converted to percentiles relative to industry peers. Four imputation techniques handle missing data, and four oversampling methods address class imbalance. Six machine learning models are trained using these imputed and balanced datasets. SHAP values are then used to interpret the models, highlighting key variables. Each step of this framework is detailed in the following subsections.

\subsection{Data Collection}
In this study, we analyze the impact of 46 independent variables related to corporate characteristics on the likelihood of a company being targeted by activist investors. These variables are categorized as follows: 10 governance indicators, 3 ownership metrics, 6 technical indicators, 7 return measures, 10 valuation metrics, and 10 operational indicators, with details presented in Table \ref{tab:overview}.

\begin{figure}
\includegraphics[width=\textwidth]{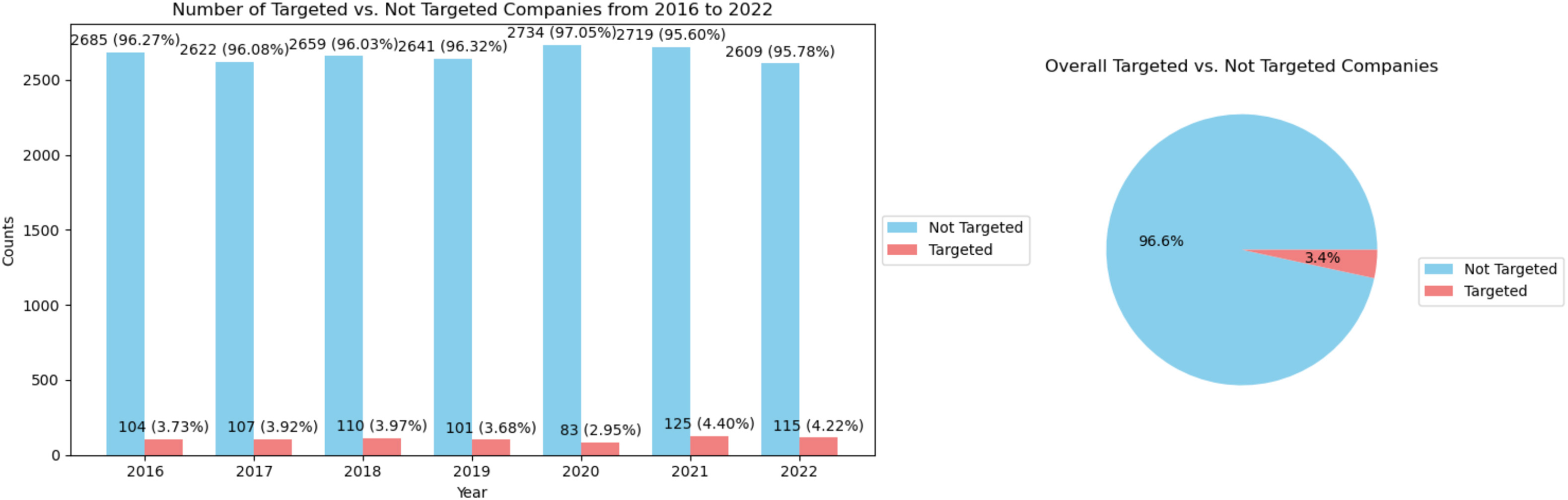}
\caption{Distributions of the targeted and non-targeted companies in the dataset. From 2016 to 2022, around 2\% to 4\% of the companies in Russell 3000 were targeted by activist funds, with the mean value of 3.4\% in the entire period.} \label{target proportion}
\end{figure}

The independent variables, sourced from the Bloomberg Terminal, comprises year-end snapshots for companies listed on the Russell 3000 index from 2016 to 2022. We chose the Russell 3000 index as the subject of this study because it represents approximately 98\% of the investable U.S. equity market, covering both large-cap and small- to mid-cap companies \footnote[1]{\url{https://research.ftserussell.com/Analytics/FactSheets/Home/DownloadSingleIssue?issueName=US3000USD&isManual=True}}. Additionally, the U.S. market provides the most comprehensive data on corporate characteristics, and is the market where activist campaigns are the most active, making it an ideal setting for our analysis. For the dependent variable that indicates if a company is targeted or not, we utilize data on activist campaigns reported from 2011 to 2023, also obtained from the Bloomberg Terminal. A company is considered a target if it becomes the subject of an activist campaign within the subsequent 12 months. Firms actively under a campaign are temporarily excluded from the dataset and reinstated after the campaign concludes. Our final dataset consists of 19,414 instances, with approximately 3.4\% of these companies being identified as targets (see Fig. \ref{target proportion}).

\subsection{Data Preprocessing}
\subsubsection{Conversion to Ratio Values \& Industry Grouping}

Some non-ratio values, like capital expenditure, working capital, total compensations to executives, are colinear with the size of the companies. It is therefore inappropriate to directly compare these figures. To account for this, we transform all non-ratio variables into ratio equivalents.

For the operations and valuation categories, values are transformed into percentiles, segmented by industry and year according to the Bloomberg Industry Classification Standard (BICS)\footnote[2]{\url{https://assets.bbhub.io/professional/sites/10/BICS-2024-Changes.pdf}}. We apply BICS Level 3 grouping when a category contains 10 or more instances. If fewer, we default to BICS Level 2 for classification.This approach allows for straightforward comparisons across different industries.

\subsubsection{Data Imputation}

A considerable portion of dataset contains missing values, posing challenges to analysis (see Fig. \ref{missing}). Imputation methods estimated and fill in these gaps, enabling robust analysis without discarding incomplete observations. We use four different methods: Mean Imputation, K-nearest neighbors (KNN) imputations (k = 5) \citep{batista2002study}, Multivariate Imputation by Chained Equations using LightGBM (MiceForest) \citep{stekhoven2015missforest}, and Generative Adversarial Imputation Networks (GAIN) \citep{yoon2018gain}. For further details of the methods, refer to \hyperref[imputationmethods]{Appendix B}.

For all imputation methods except for mean imputation, we perform the imputation process within the same category of variables, incorporating the one-hot encoded values of the companies' year. This approach ensures that imputations are conducted among variables that share predictive relationships and also accounts for overarching yearly trends. To prevent target leakage, we apply these data imputation techniques only to the training set. For the test set, we exclusively use median imputation.

\subsubsection{Data Oversampling}
In addressing the challenge posed by the dataset, where only 3.4\% of the entries are labeled as targeted, we confronted a significant imbalance between the majority and minority classes. Highly imbalanced data presents an additional challenge, as most models tend to favor the majority class, and in extreme situations, they may completely overlook the minority class. To mitigate this issue and enhance the robustness of our analysis, we conduct oversampling to adjust the ratio of targeted and non-targeted entries to 50:50. 

Specifically, we implement 4 oversampling techniques. These include Random Oversampling Example (ROSE) \citep{zheng2015oversampling}, Synthetic Minority Oversampling Technique (SMOTE) \citep{chawla2002smote}, borderline SMOTE, and Adaptive Synthetic Sampling (ADASYN) \citep{he2008adasyn}. For further details of the methods, refer to \hyperref[oversamplingsection]{Appendix C}.

\begin{table*}[!htbp] 
\centering
\scriptsize
\begin{tabular}{>{\raggedright\arraybackslash}p{2cm} >{\raggedright\arraybackslash}p{5cm} >{\raggedright\arraybackslash}p{10cm}}
\toprule
\textbf{Category} & \textbf{Variable} & \textbf{Explanation} \\
\midrule
Governance & Existence of dual-class voting rights & Dual-class voting hinders activist victory in proxy fights.\\
 & CEO's tenure & CEO changes suggest company dissatisfaction; new CEOs may be more vulnerable.\\
 & CEO is female & A company with a female CEO is more likely to be targeted.\\
 & Board size & Less effort needed to gain support from the smaller-sized board.\\
 & Existence of classified board system & Classified boards increase takeover costs by preventing full board replacement in one election. \\
 & Existence of poison pill & Poison pills raise takeover costs. \\
 & Buyback yield & Low repurchases may reflect inadequate shareholder value return. \\
 & Dividend payout ratio & A low dividend payout ratio could indicate insufficient value return to shareholders. \\
 & Free cash flow to total compensations to executives & Comparatively high executives compensation points to management inefficiency.\\
 & Free cash flow to total compensations to board members & Comparatively high board compensation points to management inefficiency. \\
\midrule
Ownership 
 & Free float percentage & Non-float shares typically align with management. \\
 & Percentage of institutional ownership & High institutional ownership may indicate that there are fewer entities for activists to pursue to launch a campaign. \\
 & Percentage of insider ownership & Low insider ownership may suggest that company executives and directors have a limited stake in the company's success \\
\midrule
Technical 
 & 30-day average trading volume to outstanding shares & High volume allows activists to acquire shares without moving the price. \\
 & 14-day Relative Strength Index & Low RSI indicates stock is oversold; provides favorable entry points. \\
 & 30-day Relative Strength Index & Low RSI indicates stock is oversold; provides favorable entry points.\\
 & 30-day volatility & High volatility can provide activist funds with favorable entry points. \\
 & 90-day volatility &  High volatility can provide activist funds with favorable entry points.\\
 & 180-day volatility &  High volatility can provide activist funds with favorable entry points.\\
\midrule
Return & 5-year total return & Long-term negative return may indicate shareholder dissatisfaction. \\
 & 4-year total return &  Long-term negative return may indicate shareholder dissatisfaction.\\
 & 3-year total return &  Long-term negative return may indicate shareholder dissatisfaction.\\
 & 2-year total return &  Long-term negative return may indicate shareholder dissatisfaction.\\
 & 1-year total return &  Short-term negative return may indicate shareholder dissatisfaction.\\
 & 6-month total return & Short-term negative return may indicate shareholder dissatisfaction. \\
 & 3-month total return & Short-term negative return may indicate shareholder dissatisfaction.\\
\midrule
Valuation & Return on equity (ROE) & Low ROE may indicate inefficiency such that it never returns satisfactory returns to the shareholders. \\
 & Return on invested capital (ROIC) & Low ROIC may indicate inefficient use of invested capital. \\
 & Assets to equity & Low asset to equity may signal excessive leverage. \\
 & Earnings per share (EPS) & Low EPS may indicate inefficiency or low profitability. \\
 & Price-earnings ratio (PER) & Low valuation makes a company more attractive to activists. \\
 & Enterprise value to sales &  Low valuation makes a company more attractive to activists.\\
 & Tobin's Q ratio &  Low valuation makes a company more attractive to activists.\\
 & Price-to-book ratio (PBR) & Low valuation makes a company more attractive to activists.\\
 & EV/EBITDA & Low valuation makes a company more attractive to activists.\\
 & Enterprise value to asset & Low valuation makes a company more attractive to activists.\\
\midrule
Operation & Free cash flow to capex & Excessive capex may indicate inefficiency, eroding shareholder value, or misaligned with shareholder interests. \\
& Current ratio & Excessive working capital ratio indicates inefficiency and low liquidity. \\
& EBITDA margin & Low margin indicates the possibility of improvement. \\
& Sales to total assets & Sales leverage is an indicator of how efficiently companies use assets. \\
& Employee growth rate & High growth rate of employees may indicate excessively aggressive expansion. \\
& Free cash flow yield & Excessively high FCF yield may signal insufficient use of cash. \\
& Sales growth rate & Low sales growth rate may indicate stagnation of a company. \\
& Interest coverage ratio & Low interest coverage ratio may indicate potential risk of insolvency. \\
& Cash conversion cycle & High cash conversion cycle might indicate liquidity problem \\
& Net debt to EBITDA & High net debt compared to EBITDA may indicate excessive debt level. \\
\bottomrule
\end{tabular}
\caption{Overview of 46 variables. Conversion to percentile compared to industry peers in the same year is done for valuation and operation variables }

\label{tab:overview}
\end{table*}

\begin{figure}
\centering
\includegraphics[width=0.5\textwidth]{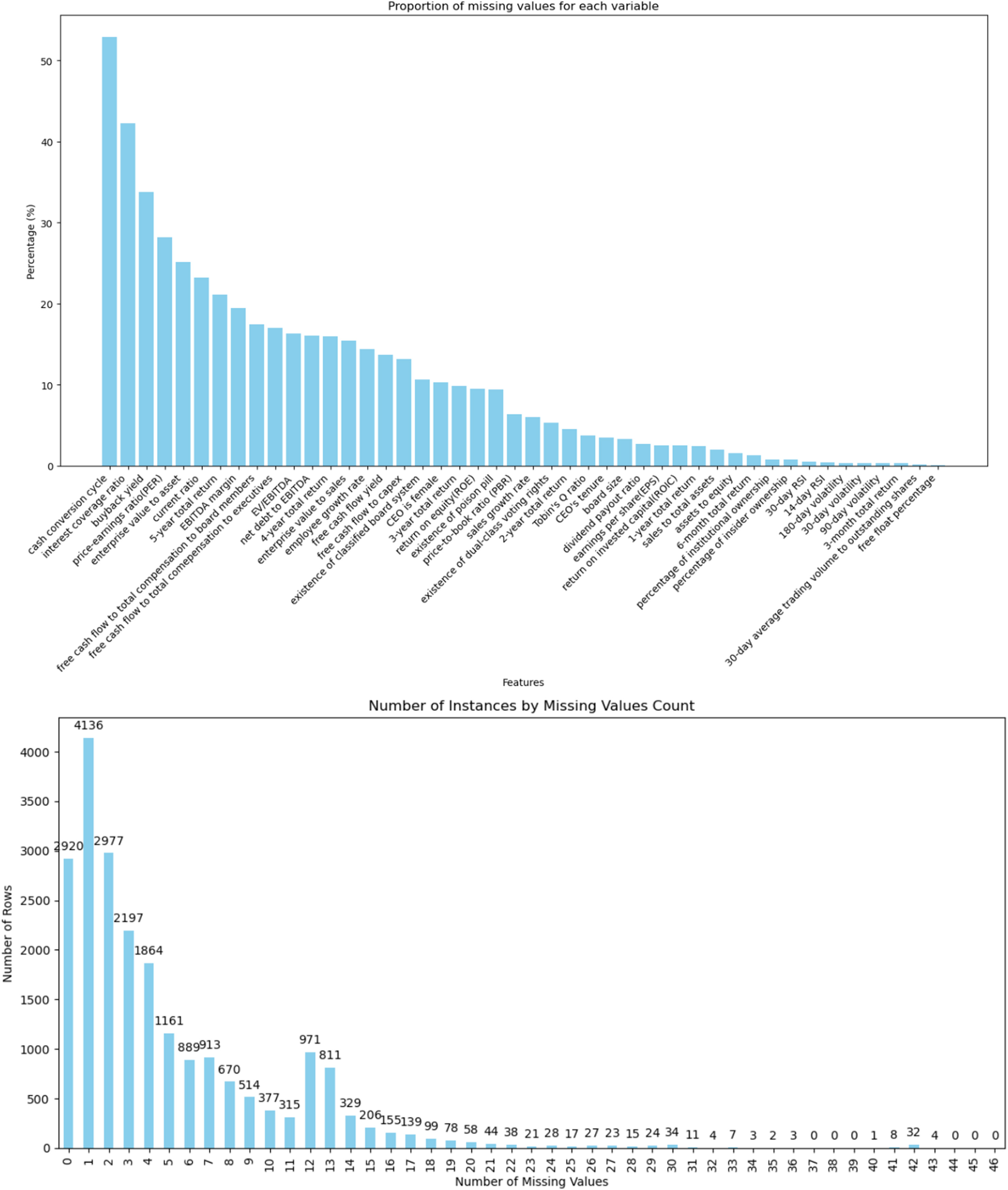}
\caption{The first subfigure shows the proportion of missing values for each variable.The second subfigure shows the number of instances with different number of missing values for 46 variables.} \label{missing}
\end{figure}

\subsection{Model Training \& Metrics}
We employ six diverse machine learning models in our analysis: Logistic Regression \citep{lavalley2008logistic}, Random Forest \citep{rigatti2017random}, XGBoost \citep{chen2015xgboost}, LightGBM \citep{ke2017lightgbm}, CatBoost \citep{prokhorenkova2018catboost}, and neural networks \citep{krogh2008artificial}. For further details of the methods, refer to \hyperref[imputation]{Appendix D}.

The data is divided, allocating 80\% (15,531 instances) for training—this figure increases when oversampling is applied—and the remaining 20\% (3,883 instances) for testing. By integrating 4 imputation methods, 5 oversampling strategies (including the option of not oversampling), and 6 machine learning models, we arrive at 120 distinct configurations. Additionally, recognizing the capabilities of LightGBM, XGBoost, and CatBoost to efficiently handle sparse data, we account for these as three extra models using original data without imputation, increasing our total to 123 models.

Considering the significant imbalance within the dataset, we select the AUC-ROC (Area Under the Receiver Operating Characteristic Curve) as our principal performance metric. This choice provides a detailed assessment of a model's predictive capability, offering insight into its performance across various threshold settings and ensuring robustness against the challenges posed by the dataset's imbalance.

\subsection{Model Interpretation Method}

The trade-off between model complexity and interpretability is a key challenge in machine learning. Neural networks and various ensemble models often outperform simpler models like linear regression but fall short in interpretability. This has led to the development of various methodologies. 

In our study, we employ SHAP (SHapley Additive exPlanations) to analyze and interpret the contributions of different features in our models. Grounded in cooperative game theory, SHAP provided a robust framework for explaining the output of any machine learning model \citep{LundbergLee2017}. We choose this method primarily for two reasons. First, the effectiveness of SHAP in highlighting the relative importance of variables was extensively validated in previous research. Second, unlike model-specific tools such as coefficients in linear regression or MDI (mean decrease in impurity) and the number of splits in tree-based models like Random Forest or LightGBM, SHAP is model-agnostic. This universality makes it a versatile tool for application across various model architectures.

\section{Results}
\subsection{Model Performance}
In an evaluation of 123 different combinations of imputation, oversampling, and machine learning methods, 11 models achieved an AUC-ROC score above 0.7, with 4 surpassing 0.75 on the test set (see Table \ref{tab:AUROC table}). The highest AUC-ROC score recorded was 0.782, attained using a combination of KNN imputation, Borderline SMOTE oversampling, and logistic regression.

Among the 11 top-performing models, 10 utilized machine learning-based imputation, with only one employing naive imputation. None of the three models capable of handling sparse data surpassed an AUC-ROC of 0.7. The result suggests that machine learning-based imputation methods significantly enhanced predictive performance.

A similar trend was observed with oversampling methods, as only one model never employed oversampling. This suggests that oversampling effectively addressed the class imbalance problem, which can often cause models to overlook or undervalue the characteristics of the minority class. Notably, all 10 oversampling techniques used in the top 11 models employed Borderline SMOTE. Borderline SMOTE targets instances near the decision boundary between the two classes. The effectiveness of Borderline SMOTE indicates the possibility of a distinct boundary between targeted and non-targeted instances in the vector space, a subject that could be further explored in future research.

In terms of machine learning methods, all but neural networks appeared among the top 11 models. All of the 4 models that surpassed an AUC-ROC score of 0.75 employed traditional logistic regression, indicating that more advanced machine learning models, such as neural networks and ensemble methods, did not lead to superior performance. This trend, where logistic regression outperforms more complex models, has been frequently observed in many studies across various domains \citep{Wu2023}, \citep{Liew2022}, \citep{Lynam2020}, \citep{Song2021}. A possible explanation is that advanced models often require larger datasets to effectively learn the underlying patterns, while logistic regression can perform robustly even with smaller datasets. This finding emphasizes the effectiveness of logistic regression, particularly in contexts where data size is limited.

\begin{table}[ht]
\centering
\begin{tabular}{llll}
\toprule
Imputation & Oversampling & ML Method & AUC-ROC \\
\midrule
KNN & Borderline SMOTE & Logistic Regression & 0.782 \\
Median & Borderline SMOTE & Logistic Regression & 0.770 \\
GAIN & Borderline SMOTE & Logistic Regression & 0.767 \\
MiceForest & Borderline SMOTE & Logistic Regression & 0.752 \\
MiceForest & Borderline SMOTE & Random Forest & 0.727 \\

GAIN & Borderline SMOTE & Random Forest & 0.709 \\
GAIN & Borderline SMOTE & Light GBM & 0.707 \\
KNN & Borderline SMOTE & XGBoost & 0.706 \\
KNN & Borderline SMOTE & Random Forest & 0.706 \\
KNN & No Oversampling & CatBoost & 0.701\\
MiceForest & Borderline SMOTE & CatBoost & 0.700 \\

\bottomrule
\end{tabular}
\caption{Summary of 11 models that have AUC-ROC (test set) of greater than 0.7}
\label{tab:AUROC table}
\end{table}

\subsection{Model Interpretation}
\subsubsection{SHAP Value Analysis of the Best-performing Model}
\begin{figure}
\includegraphics[width=\textwidth]{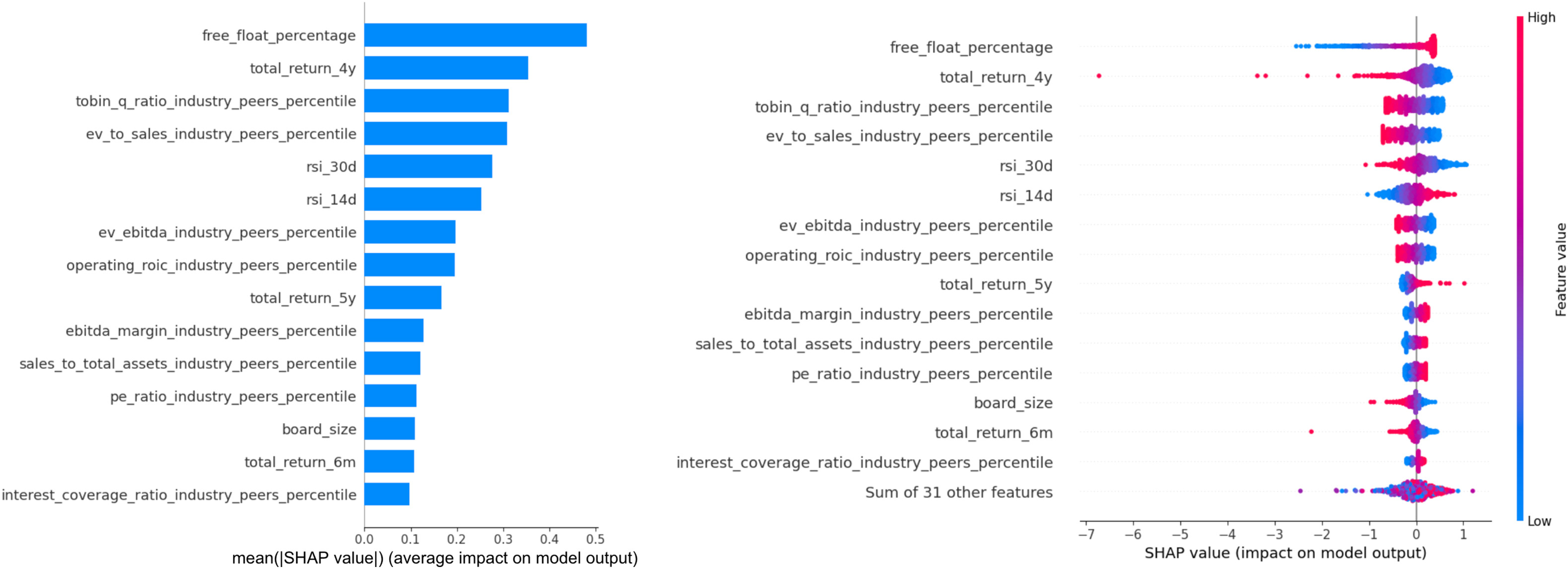}
\caption{The first subplot displays the mean absolute SHAP values of the top 15 variables for a logistic regression model trained on data processed with KNN imputation and Borderline SMOTE are shown in the first subplot. The second beeswarm subplot displays individual SHAP values for the top 15 variables across all instances.} \label{SHAP_original}
\end{figure}

SHAP values for our best-performing model, which used KNN imputation, Borderline SMOTE oversampling, and logistic regression, are presented in Fig. \ref{SHAP_original}. The top 15 most influential variables, ranked by their mean absolute SHAP values, represent a diverse mix across all categories.

Among the three ownership metrics, only the free float percentage was included in the top 15, but it had the highest mean absolute SHAP value. As illustrated in the beeswarm plot in Fig. \ref{SHAP_original}, a high free float percentage dramatically reduces the probability of being targeted by activist funds. However, having a high value does not significantly increase the likelihood of being targeted. This suggests that while a sufficiently high level of freely tradable shares does not greatly influence the target selection process, once the free float percentage falls below a certain threshold, it becomes a significant barrier for activist funds to acquire a stake, sharply decreasing the company’s appeal as a target.

Three of the seven return metrics assessed, specifically the 5-year, 4-year, and 6-month returns, made it into the top 15. The 4-year return had the highest mean absolute SHAP value, exceeding 0.3, while the one of 6-month return was slightly above 0.1, indicating that activist funds prioritize long-term performance over short-term. This suggests that companies showing poor returns over an extended period are perceived as having unrealized potential, making them preferred targets. Similar to the free float percentage, the 4-year return's beeswarm plot shows a skewed distribution: returns above a certain threshold have little impact, but below this threshold, the likelihood of being targeted increases sharply. A similar trend is observed for the 6-month return. The 5-year return yielded a counterintuitive result: higher returns were associated with a greater likelihood of being targeted. Given the strong positive correlation between 4-year and 5-year returns, this result is particularly surprising. To verify its reliability, we performed a robustness check on the second, third, and fourth best-performing models (Fig. \ref{robustness-check-1}). In none of these models did the 5-year return rank as a key variable, suggesting that the result may require further scrutiny. One possible explanation could be the missing data. As shown in Fig. \label{missing}, while 4-year returns were missing for around 15\% of the data, 5-year returns had a higher missing rate of 21\%. This may have distorted the results, though further research is needed to confirm this hypothesis.

Among traditional valuation metrics, Tobin Q ratio, EV to sales, and EV/EBITDA, also appeared in the top 15 variables. These metrics showed intuitive results: companies were more likely to be targeted when undervalued and less likely when overvalued. Due to the high correlation between these valuation metrics, a consistent pattern emerged. The beeswarm plots showed uniform distribution, indicating a linear relationship between these metrics and the likelihood of being targeted. This likely stems from the fact that the metrics were converted to percentiles relative to industry peers using BICS classification, meaning the more undervalued a company is compared to its peers, the higher the probability of being targeted increases in a linear fashion.

Among the operational metrics used to assess a company's profitability and financial efficiency, Operating ROIC, EBITDA margin, sales to total assets, and interest coverage ratio appeared in the top 15. Like the valuation metrics, these operational metrics were converted into percentiles relative to industry peers, leading to uniformly distributed beeswarm plots. However, they exhibited opposing trends: a lower Operating ROIC increased the likelihood of being targeted, suggesting that activist funds focus on companies with lower operational efficiency, while the other three metrics indicated that firms with stronger operational performance were more likely to be targeted. This suggests two possibilities: activist funds may target both inefficient companies for potential operational improvements and well-performing but undervalued firms. Alternatively, the results may lack reliability, as none of these metrics had a mean absolute SHAP value above 0.2, and they were occasionally absent from the top 15 when tested with other models in the robustness check (Fig. \ref{robustness-check-1}). Further research is needed to better understand which operational characteristics activist funds prioritize.

Among the technical indicators, the 30-day RSI and 14-day RSI, both signaling overbought/oversold conditions, exceeded a mean absolute SHAP value of 0.2 and appeared in all three robustness check models. Interestingly, the 30-day RSI showed that the more oversold a stock was, the higher the likelihood of being targeted, while the 14-day RSI showed the opposite trend. One interpretation is that activist funds start buying when a stock is oversold, raising the RSI over time. However, this remains speculative, and further research is needed.

In terms of governance, only board size appeared in the top 15 variables. Smaller boards were more likely to be targeted, potentially because fewer board members make it easier for activist funds to influence decisions within the company.

\subsubsection{Robustness Check}
\begin{figure}
\centering
\includegraphics[width=\textwidth]{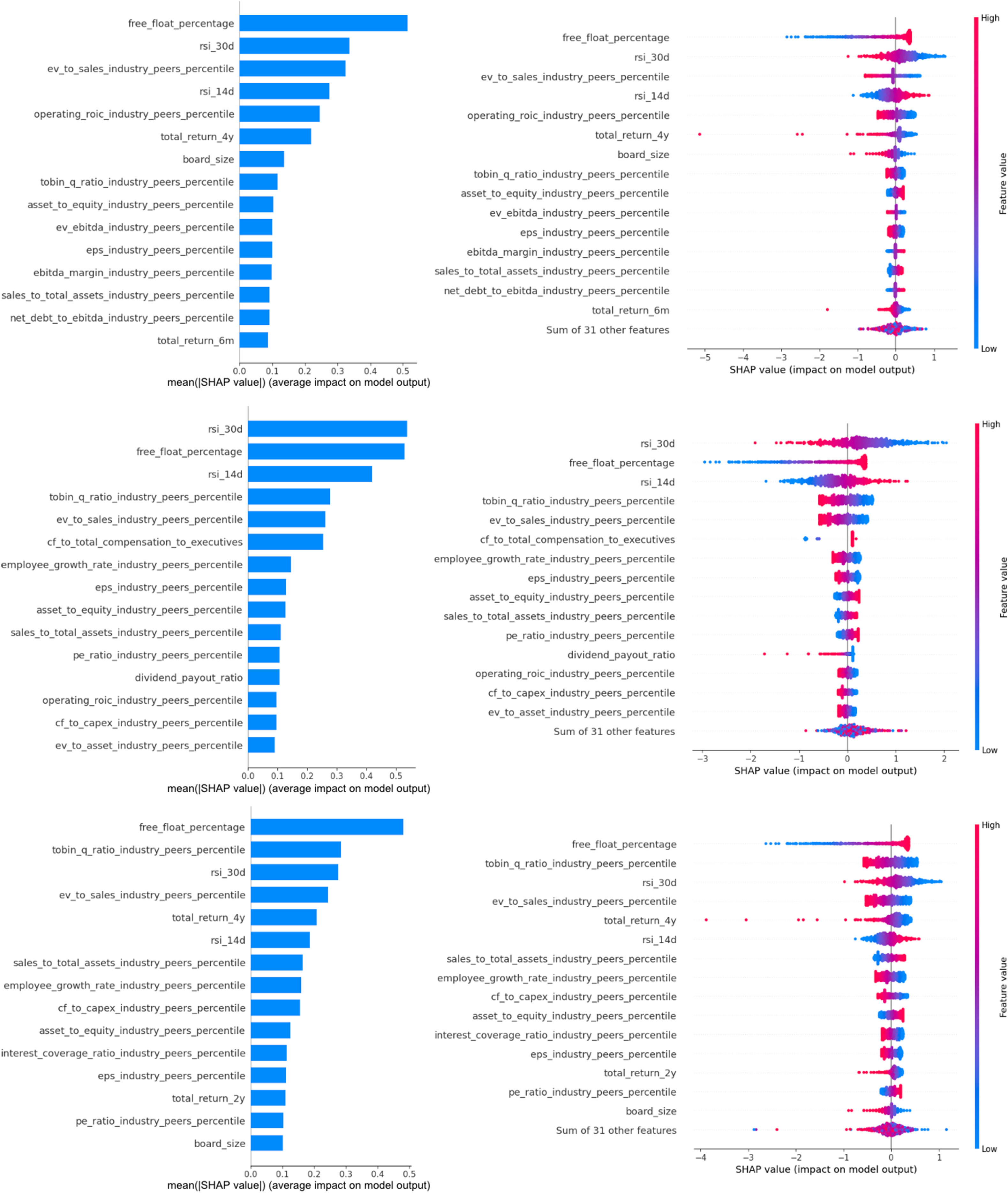}
\caption{Summary plots of SHAP values for three models with AUC-ROC scores above 0.75. The first row features median imputation with Borderline SMOTE and logistic regression; the second row, GAIN imputation with Borderline SMOTE and logistic regression; and the third row, MiceForest imputation with Borderline SMOTE and logistic regression.} \label{robustness-check-1}
\end{figure}

\begin{figure}
\centering
\includegraphics[width=0.7\textwidth]{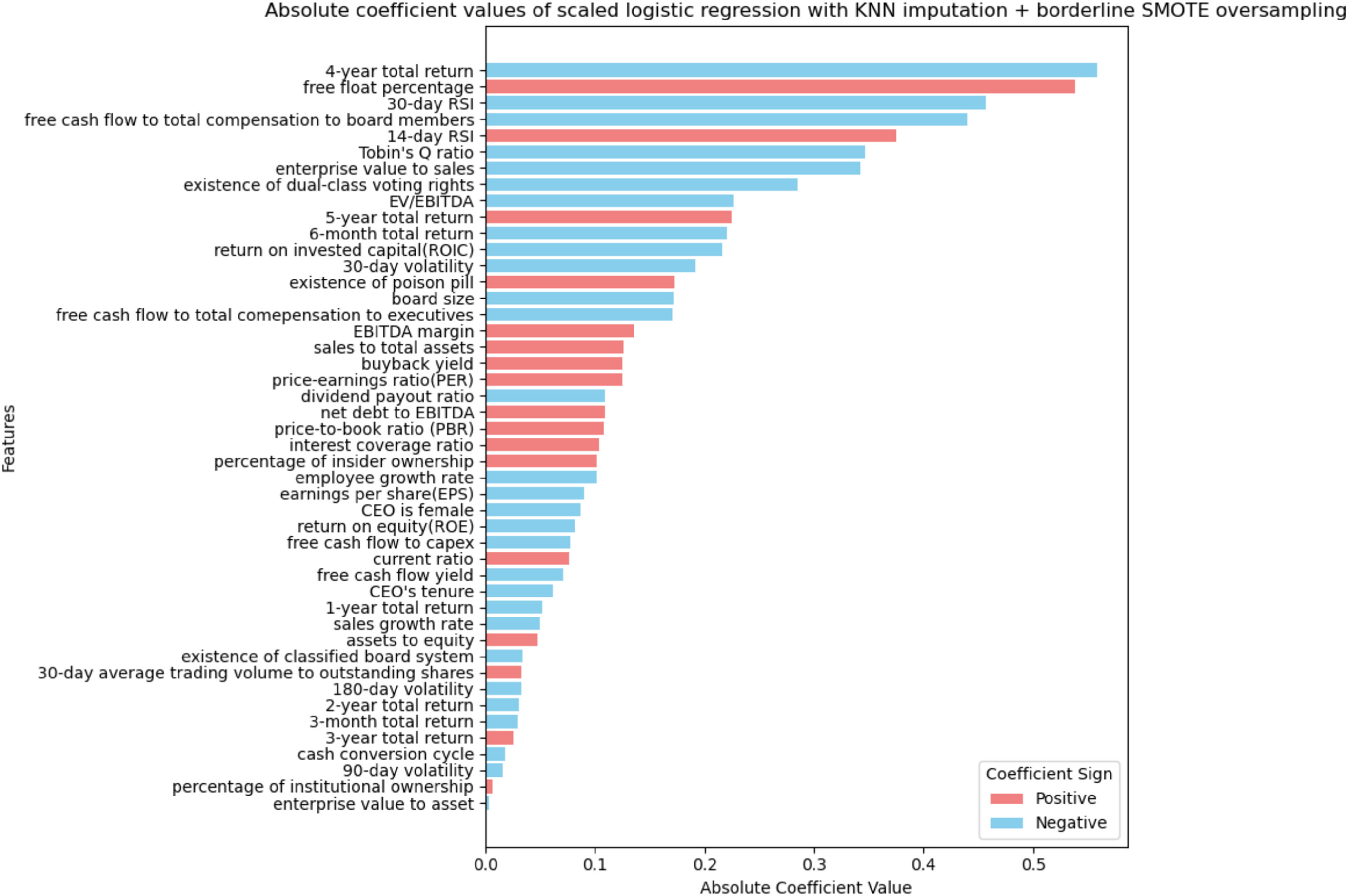}
\caption{Coefficients in the scaled logistic regression of the best-performing model (KNN imputation + Borderline SMOTE + Logistic Regression)} \label{robustness-check-2}
\end{figure}

To ensure robustness, we evaluated the Shapley values of three other top-performing models, each with AUC-ROC scores above 0.75 (refer to Fig. \ref{robustness-check-1}). The results confirmed consistency in the variable rankings across the models, with only minor changes in ranks. Some governance-related variables such as the board size and dividend payout ratio. Additionally, given that our models utilized logistic regression, we examined the scaled coefficients to further validate robustness refer to Fig. \ref{robustness-check-2}. While most variable rankings remained consistent, governance-related variables such as the ratio of free cash flow to compensation for board members and the presence of dual-class voting rights were ranked highly. These differences could be attributed to various factors, including multicollinearity among variables and the presence of outliers. These findings open up opportunities for a deeper understanding of governance-related factors in target selection, highlighting the need for further investigation into the underlying reasons.

\section{Conclusion}
In this paper, we develop a machine-learning model to predict potential targets of activist funds. Initially, we conduct a comprehensive literature review to identify key factors that influence the likelihood of a company being targeted by activists. Based on this review, we assemble a set of 46 variables providing a detailed overview of a company's various aspects. Utilizing data from Russell 3000 companies from 2016 to 2022 and activism campaign data from 2013 to 2023, we train models using 123 different combinations of data imputation, oversampling, and machine learning methodologies. The best-performing model achieves an AUC-ROC score of 0.782. Additionally, we employ SHAP analysis to gain deeper insights into the decision-making processes of our model, revealing that a mix of variables related to valuation, operation, technical aspects, returns, and ownership are considered when activists select their targets.

This research has significant theoretical and practical implications. Theoretically, it advances our understanding by constructing a data-driven supervised machine learning model that integrates multiple company aspects and uses SHAP to identify the most influential variables. This comprehensive and systematic approach can serve as a foundation for future studies exploring additional variables. Practically, the model offers several benefits. For activists, it provides a tool to reduce the time and cost involved in screening potential targets. For companies, it enables preemptive measures by indicating the likelihood of being targeted. For other investors, it offers a chance to profit from anticipated improvements in profitability and market value following activist interventions.

\section{Future Directions}
The study presents several limitations. First, there is potential for improvement in the data structure. Our model treats multiple entries for the same company at different times as separate instances, which may overlook the unique characteristics and individual effects of different firms, thereby introducing bias. Additionally, our labeling approach categorizes a company as a target only if it becomes a target within the next 12 months, regardless of the exact date the campaign begins. This approach might distort or fail to capture the significance of short-term characteristics, as campaigns are initiated at various times throughout the year. It may also fail to identify companies that exhibit sufficient characteristics of being targeted but were not actually targeted within the study's timeframe. Therefore, an alternative approach that addresses these issues needs to be explored in the future.

Second, while machine learning models demonstrate strong predictive power and SHAP helps clarify the influence of independent variables, these methods reveal correlations rather than causality. Especially, some counterintuitive results in the SHAP analysis are challenging to interpret. Developing a more rigorous causal model to better capture the mechanisms driving activists' decision-making in target selection could provide deeper insights into these complexities.

Additionally, future research could expand the scope of our study by incorporating time-trend data to observe changes in profitability over time. Another promising avenue would involve analyzing the behavior of activist funds in financial markets outside the US public market, to better understand inter-market heterogeneity and dynamics. This broader perspective could provide deeper insights into the strategies and impacts of activist investments globally.

\section*{Conflicts of Interest}
The authors declare no conflict of interest.

\section*{Acknowledgement}
The research wishes to acknowledge Mr. Adam Kommel of Bloomberg LP for his feedback and input surrounding the research and its development.

\newpage

%
%
%
%
\bibliographystyle{elsarticle-harv}\biboptions{authoryear}
\bibliography{bibliographyfile}

\begin{thebibliography}{33}
\expandafter\ifx\csname natexlab\endcsname\relax\def\natexlab#1{#1}\fi
\providecommand{\url}[1]{\texttt{#1}}
\providecommand{\href}[2]{#2}
\providecommand{\path}[1]{#1}
\providecommand{\DOIprefix}{doi:}
\providecommand{\ArXivprefix}{arXiv:}
\providecommand{\URLprefix}{URL: }
\providecommand{\Pubmedprefix}{pmid:}
\providecommand{\doi}[1]{\href{http://dx.doi.org/#1}{\path{#1}}}
\providecommand{\Pubmed}[1]{\href{pmid:#1}{\path{#1}}}
\providecommand{\bibinfo}[2]{#2}
\ifx\xfnm\relax \def\xfnm[#1]{\unskip,\space#1}\fi
\bibitem[{Albuquerque et~al.(2019)Albuquerque, Durnev and Koskinen}]{Albuquerque2019}
\bibinfo{author}{Albuquerque, R.}, \bibinfo{author}{Durnev, A.}, \bibinfo{author}{Koskinen, Y.}, \bibinfo{year}{2019}.
\newblock \bibinfo{title}{Corporate social responsibility and firm risk: Theory and evidence}.
\newblock \bibinfo{journal}{Management science} .
\bibitem[{Aluchna(2023)}]{Aluchna2023}
\bibinfo{author}{Aluchna, M.}, \bibinfo{year}{2023}.
\newblock \bibinfo{title}{Agency theory}, in: \bibinfo{booktitle}{Encyclopedia of Sustainable Management}. \bibinfo{publisher}{Springer}, pp. \bibinfo{pages}{87--95}.
\bibitem[{Bancel and Mittoo(2014)}]{BancelMittoo2014}
\bibinfo{author}{Bancel, F.}, \bibinfo{author}{Mittoo, U.R.}, \bibinfo{year}{2014}.
\newblock \bibinfo{title}{The gap between the theory and practice of corporate valuation: Survey of european experts}.
\newblock \bibinfo{journal}{Journal of Applied Corporate Finance} \bibinfo{volume}{26}, \bibinfo{pages}{106--117}.
\bibitem[{Barko et~al.(2022)Barko, Cremers and Renneboog}]{Barko2022}
\bibinfo{author}{Barko, T.}, \bibinfo{author}{Cremers, M.}, \bibinfo{author}{Renneboog, L.}, \bibinfo{year}{2022}.
\newblock \bibinfo{title}{Shareholder engagement on environmental, social, and governance performance}.
\newblock \bibinfo{journal}{Journal of Business Ethics} \bibinfo{volume}{180}, \bibinfo{pages}{777--812}.
\bibitem[{Batista et~al.(2002)Batista, Monard et~al.}]{batista2002study}
\bibinfo{author}{Batista, G.E.}, \bibinfo{author}{Monard, M.C.}, et~al., \bibinfo{year}{2002}.
\newblock \bibinfo{title}{A study of k-nearest neighbour as an imputation method.}
\newblock \bibinfo{journal}{His} \bibinfo{volume}{87}, \bibinfo{pages}{48}.
\bibitem[{Bebchuk et~al.(2015)Bebchuk, Brav and Jiang}]{Bebchuk2015}
\bibinfo{author}{Bebchuk, L.A.}, \bibinfo{author}{Brav, A.}, \bibinfo{author}{Jiang, W.}, \bibinfo{year}{2015}.
\newblock \bibinfo{title}{The long-term effects of hedge fund activism}.
\newblock \bibinfo{type}{Technical Report}. National Bureau of Economic Research.
\bibitem[{Boyson and Mooradian(2011)}]{BoysonMooradian2011}
\bibinfo{author}{Boyson, N.M.}, \bibinfo{author}{Mooradian, R.M.}, \bibinfo{year}{2011}.
\newblock \bibinfo{title}{Corporate governance and hedge fund activism}.
\newblock \bibinfo{journal}{Review of Derivatives Research} \bibinfo{volume}{14}, \bibinfo{pages}{169--204}.
\bibitem[{Brav et~al.(2008)Brav, Jiang, Partnoy and Thomas}]{BravJiangPartnoyThomas2008}
\bibinfo{author}{Brav, A.}, \bibinfo{author}{Jiang, W.}, \bibinfo{author}{Partnoy, F.}, \bibinfo{author}{Thomas, R.}, \bibinfo{year}{2008}.
\newblock \bibinfo{title}{Hedge fund activism, corporate governance, and firm performance}.
\newblock \bibinfo{journal}{The Journal of Finance} \bibinfo{volume}{63}, \bibinfo{pages}{1729--1775}.
\bibitem[{Briggs(2006)}]{Briggs2007}
\bibinfo{author}{Briggs, T.W.}, \bibinfo{year}{2006}.
\newblock \bibinfo{title}{Corporate governance and the new hedge fund activism: An empirical analysis}.
\newblock \bibinfo{journal}{J. Corp. L.} \bibinfo{volume}{32}, \bibinfo{pages}{681}.
\bibitem[{Chawla et~al.(2002)Chawla, Bowyer, Hall and Kegelmeyer}]{chawla2002smote}
\bibinfo{author}{Chawla, N.V.}, \bibinfo{author}{Bowyer, K.W.}, \bibinfo{author}{Hall, L.O.}, \bibinfo{author}{Kegelmeyer, W.P.}, \bibinfo{year}{2002}.
\newblock \bibinfo{title}{Smote: synthetic minority over-sampling technique}.
\newblock \bibinfo{journal}{Journal of artificial intelligence research} \bibinfo{volume}{16}, \bibinfo{pages}{321--357}.
\bibitem[{Chen et~al.(2015)Chen, He, Benesty, Khotilovich, Tang, Cho, Chen, Mitchell, Cano, Zhou et~al.}]{chen2015xgboost}
\bibinfo{author}{Chen, T.}, \bibinfo{author}{He, T.}, \bibinfo{author}{Benesty, M.}, \bibinfo{author}{Khotilovich, V.}, \bibinfo{author}{Tang, Y.}, \bibinfo{author}{Cho, H.}, \bibinfo{author}{Chen, K.}, \bibinfo{author}{Mitchell, R.}, \bibinfo{author}{Cano, I.}, \bibinfo{author}{Zhou, T.}, et~al., \bibinfo{year}{2015}.
\newblock \bibinfo{title}{Xgboost: extreme gradient boosting}.
\newblock \bibinfo{journal}{R package version 0.4-2} \bibinfo{volume}{1}, \bibinfo{pages}{1--4}.
\bibitem[{DesJardine and Durand(2020)}]{DesJardineDurand2020}
\bibinfo{author}{DesJardine, M.R.}, \bibinfo{author}{Durand, R.}, \bibinfo{year}{2020}.
\newblock \bibinfo{title}{Disentangling the effects of hedge fund activism on firm financial and social performance}.
\newblock \bibinfo{journal}{Strategic Management Journal} \bibinfo{volume}{41}, \bibinfo{pages}{1054--1082}.
\newblock \URLprefix \url{https://doi.org/10.1002/smj.3126}, \DOIprefix\doi{10.1002/smj.3126}.
\bibitem[{Flammer(2015)}]{Flammer2015}
\bibinfo{author}{Flammer, C.}, \bibinfo{year}{2015}.
\newblock \bibinfo{title}{Does corporate social responsibility lead to superior financial performance? a regression discontinuity approach}.
\newblock \bibinfo{journal}{Management science} \bibinfo{volume}{61}, \bibinfo{pages}{2549--2568}.
\bibitem[{Francis et~al.(2021)Francis, Hasan, Shen and Wu}]{Francis2021}
\bibinfo{author}{Francis, B.B.}, \bibinfo{author}{Hasan, I.}, \bibinfo{author}{Shen, Y.V.}, \bibinfo{author}{Wu, Q.}, \bibinfo{year}{2021}.
\newblock \bibinfo{title}{Do activist hedge funds target female ceos? the role of ceo gender in hedge fund activism}.
\newblock \bibinfo{journal}{Journal of Financial Economics} \bibinfo{volume}{141}, \bibinfo{pages}{372--393}.
\bibitem[{He et~al.(2008)He, Bai, Garcia and Li}]{he2008adasyn}
\bibinfo{author}{He, H.}, \bibinfo{author}{Bai, Y.}, \bibinfo{author}{Garcia, E.A.}, \bibinfo{author}{Li, S.}, \bibinfo{year}{2008}.
\newblock \bibinfo{title}{Adasyn: Adaptive synthetic sampling approach for imbalanced learning}, in: \bibinfo{booktitle}{2008 IEEE international joint conference on neural networks (IEEE world congress on computational intelligence)}, \bibinfo{organization}{Ieee}. pp. \bibinfo{pages}{1322--1328}.
\bibitem[{Ke et~al.(2017)Ke, Meng, Finley, Wang, Chen, Ma, Ye and Liu}]{ke2017lightgbm}
\bibinfo{author}{Ke, G.}, \bibinfo{author}{Meng, Q.}, \bibinfo{author}{Finley, T.}, \bibinfo{author}{Wang, T.}, \bibinfo{author}{Chen, W.}, \bibinfo{author}{Ma, W.}, \bibinfo{author}{Ye, Q.}, \bibinfo{author}{Liu, T.Y.}, \bibinfo{year}{2017}.
\newblock \bibinfo{title}{Lightgbm: A highly efficient gradient boosting decision tree}.
\newblock \bibinfo{journal}{Advances in neural information processing systems} \bibinfo{volume}{30}.
\bibitem[{Krogh(2008)}]{krogh2008artificial}
\bibinfo{author}{Krogh, A.}, \bibinfo{year}{2008}.
\newblock \bibinfo{title}{What are artificial neural networks?}
\newblock \bibinfo{journal}{Nature biotechnology} \bibinfo{volume}{26}, \bibinfo{pages}{195--197}.
\bibitem[{LaValley(2008)}]{lavalley2008logistic}
\bibinfo{author}{LaValley, M.P.}, \bibinfo{year}{2008}.
\newblock \bibinfo{title}{Logistic regression}.
\newblock \bibinfo{journal}{Circulation} \bibinfo{volume}{117}, \bibinfo{pages}{2395--2399}.
\bibitem[{Lazard(2018)}]{Lazard2018}
\bibinfo{author}{Lazard}, \bibinfo{year}{2018}.
\newblock \bibinfo{title}{{ Annual Review of Shareholder Activism}}.
\newblock \bibinfo{howpublished}{\url{https://www.lazard.com/media/450655/lazards-review-of-shareholder-activism-1h-2018.pdf}}.
\newblock \bibinfo{note}{[Online;]}.
\bibitem[{Liew et~al.(2022)Liew, Kovacs, Rügamer and Royuela}]{Liew2022}
\bibinfo{author}{Liew, B.X.}, \bibinfo{author}{Kovacs, F.M.}, \bibinfo{author}{Rügamer, D.}, \bibinfo{author}{Royuela, A.}, \bibinfo{year}{2022}.
\newblock \bibinfo{title}{Machine learning versus logistic regression for prognostic modelling in individuals with non-specific neck pain}.
\newblock \bibinfo{journal}{European Spine Journal} \bibinfo{volume}{31}, \bibinfo{pages}{2082--2091}.
\newblock \DOIprefix\doi{10.1007/s00586-022-07188-w}.
\bibitem[{Lundberg and Lee(2017)}]{LundbergLee2017}
\bibinfo{author}{Lundberg, S.M.}, \bibinfo{author}{Lee, S.I.}, \bibinfo{year}{2017}.
\newblock \bibinfo{title}{A unified approach to interpreting model predictions}.
\newblock \bibinfo{journal}{Advances in neural information processing systems} \bibinfo{volume}{30}.
\bibitem[{Lynam et~al.(2020)Lynam, Dennis, Owen, Oram, Jones, Shields and Ferrat}]{Lynam2020}
\bibinfo{author}{Lynam, A.L.}, \bibinfo{author}{Dennis, J.M.}, \bibinfo{author}{Owen, K.R.}, \bibinfo{author}{Oram, R.A.}, \bibinfo{author}{Jones, A.G.}, \bibinfo{author}{Shields, B.M.}, \bibinfo{author}{Ferrat, L.A.}, \bibinfo{year}{2020}.
\newblock \bibinfo{title}{Logistic regression has similar performance to optimised machine learning algorithms in a clinical setting: application to the discrimination between type 1 and type 2 diabetes in young adults}.
\newblock \bibinfo{journal}{Diagnostic and Prognostic Research} \bibinfo{volume}{4}, \bibinfo{pages}{6}.
\newblock \DOIprefix\doi{10.1186/s41512-020-00075-2}.
\bibitem[{Mukhlynina and Nyborg(2016)}]{MukhlyninaNyborg2016}
\bibinfo{author}{Mukhlynina, L.}, \bibinfo{author}{Nyborg, K.G.}, \bibinfo{year}{2016}.
\newblock \bibinfo{title}{The choice of valuation techniques in practice: education versus profession}.
\newblock \bibinfo{journal}{Swiss Finance Institute Research Paper} .
\bibitem[{Pfirrmann and Eichner(2024)}]{PfirrmannEichner2023}
\bibinfo{author}{Pfirrmann, M.}, \bibinfo{author}{Eichner, K.}, \bibinfo{year}{2024}.
\newblock \bibinfo{title}{How activist investors value target firms: Evidence from hedge fund presentations}.
\newblock \bibinfo{journal}{Journal of Corporate Accounting \& Finance} .
\bibitem[{Prokhorenkova et~al.(2018)Prokhorenkova, Gusev, Vorobev, Dorogush and Gulin}]{prokhorenkova2018catboost}
\bibinfo{author}{Prokhorenkova, L.}, \bibinfo{author}{Gusev, G.}, \bibinfo{author}{Vorobev, A.}, \bibinfo{author}{Dorogush, A.V.}, \bibinfo{author}{Gulin, A.}, \bibinfo{year}{2018}.
\newblock \bibinfo{title}{Catboost: unbiased boosting with categorical features}.
\newblock \bibinfo{journal}{Advances in neural information processing systems} \bibinfo{volume}{31}.
\bibitem[{Rigatti(2017)}]{rigatti2017random}
\bibinfo{author}{Rigatti, S.J.}, \bibinfo{year}{2017}.
\newblock \bibinfo{title}{Random forest}.
\newblock \bibinfo{journal}{Journal of Insurance Medicine} \bibinfo{volume}{47}, \bibinfo{pages}{31--39}.
\bibitem[{Romito(2015)}]{Romito2015}
\bibinfo{author}{Romito, D.}, \bibinfo{year}{2015}.
\newblock \bibinfo{title}{{ 4 Types of Activist Investors and How to Spot Them. Harvard Business Review}}.
\newblock \bibinfo{howpublished}{\url{https://hbr.org/2015/10/4-types-of-activist-investors-and-how-to-spot-them}}.
\newblock \bibinfo{note}{[Online;]}.
\bibitem[{Song et~al.(2021)Song, Liu, Liu and Wang}]{Song2021}
\bibinfo{author}{Song, X.}, \bibinfo{author}{Liu, X.}, \bibinfo{author}{Liu, F.}, \bibinfo{author}{Wang, C.}, \bibinfo{year}{2021}.
\newblock \bibinfo{title}{Comparison of machine learning and logistic regression models in predicting acute kidney injury: A systematic review and meta-analysis}.
\newblock \bibinfo{journal}{International Journal of Medical Informatics} \bibinfo{volume}{151}, \bibinfo{pages}{104484}.
\newblock \DOIprefix\doi{10.1016/j.ijmedinf.2021.104484}.
\bibitem[{Stekhoven(2015)}]{stekhoven2015missforest}
\bibinfo{author}{Stekhoven, D.J.}, \bibinfo{year}{2015}.
\newblock \bibinfo{title}{missforest: Nonparametric missing value imputation using random forest}.
\newblock \bibinfo{journal}{Astrophysics Source Code Library} , \bibinfo{pages}{ascl--1505}.
\bibitem[{Wu et~al.(2023)Wu, Wei, Wu, Yi and Li}]{Wu2023}
\bibinfo{author}{Wu, T.}, \bibinfo{author}{Wei, Y.}, \bibinfo{author}{Wu, J.}, \bibinfo{author}{Yi, B.}, \bibinfo{author}{Li, H.}, \bibinfo{year}{2023}.
\newblock \bibinfo{title}{Logistic regression technique is comparable to complex machine learning algorithms in predicting cognitive impairment related to post intensive care syndrome}.
\newblock \bibinfo{journal}{Scientific Reports} \bibinfo{volume}{13}, \bibinfo{pages}{2485}.
\newblock \DOIprefix\doi{10.1038/s41598-023-28421-6}.
\bibitem[{Yoon et~al.(2018)Yoon, Jordon and Schaar}]{yoon2018gain}
\bibinfo{author}{Yoon, J.}, \bibinfo{author}{Jordon, J.}, \bibinfo{author}{Schaar, M.}, \bibinfo{year}{2018}.
\newblock \bibinfo{title}{Gain: Missing data imputation using generative adversarial nets}, in: \bibinfo{booktitle}{International conference on machine learning}, \bibinfo{organization}{PMLR}. pp. \bibinfo{pages}{5689--5698}.
\bibitem[{Zheng et~al.(2015)Zheng, Cai and Li}]{zheng2015oversampling}
\bibinfo{author}{Zheng, Z.}, \bibinfo{author}{Cai, Y.}, \bibinfo{author}{Li, Y.}, \bibinfo{year}{2015}.
\newblock \bibinfo{title}{Oversampling method for imbalanced classification}.
\newblock \bibinfo{journal}{Computing and Informatics} \bibinfo{volume}{34}, \bibinfo{pages}{1017--1037}.
\bibitem[{Zhu(2021)}]{Zhu2020}
\bibinfo{author}{Zhu, C.H.}, \bibinfo{year}{2021}.
\newblock \bibinfo{title}{The preventive effect of hedge fund activism: investment, ceo compensation and payout policies}.
\newblock \bibinfo{journal}{International Journal of Managerial Finance} \bibinfo{volume}{17}, \bibinfo{pages}{401--415}.

\end{thebibliography}
\newpage
\section{Appendix}
\label{imputationmethods}
\subsection*{A. Supplementary Data and Code}
Supplementary data to this research can be found at \url{https://github.com/activistprediction/activist-target-prediction}

\subsection*{B. Imputation Methods}

\subsubsection*{Mean Imputation}
Mean imputation is one of the simplest methods for handling missing data. In this approach, the missing values in a feature (or column) are replaced with the mean of the observed values for that feature. This method assumes that the missing data are random and that the mean provides a reasonable estimate of the missing entries.
\\
\\
\textbf{Mathematical Formulation:}
\begin{itemize}
    \item For each feature $X_j$ that contains missing values:
\begin{enumerate}
    \item Calculate the mean of the observed values for the feature:
    \[
    \hat{\mu}_j = \frac{1}{n_j} \sum_{i=1}^{n_j} X_{ij}
    \]
    where $n_j$ is the number of observed (non-missing) values in feature $X_j$.
    
    \item Replace all missing values $X_{ij} = \text{NA}$ in the feature with the calculated mean $\hat{\mu}_j$:
    \[
    X_{ij} = \hat{\mu}_j \quad \text{for all missing } X_{ij}.
    \]
\end{enumerate}
\end{itemize}

\subsubsection*{K-Nearest Neighbors (KNN) Imputation}
K-nearest neighbors (KNN) imputation is a technique used to fill in missing values in a dataset based on the values of its neighboring data points. The missing value is estimated by averaging or taking a weighted average of the values of the k nearest neighbors to the missing point. 
\\
\\
\textbf{Mathematical Formulation}:
\begin{itemize}
    \item Let $X \in \mathbb{R}^{n \times d}$ represent the dataset, where $X_{ij}$ is the value of feature $j$ for observation $i$, and some values are missing. 
    \item For each observation $i$, identify the $K$-nearest neighbors in the dataset by computing the distance between the rows $X_i$ and other rows, ignoring missing values in the distance calculation:

\[
d(X_i, X_k) = \sum_{j \in \text{Non-missing}}(X_{ij} - X_{kj})^2
\]
\item The missing value $X_{ij}$ is imputed by averaging the corresponding feature values of the $K$-nearest neighbors, weighted by their distances:

\[
\hat{X}_{ij} = \frac{1}{K} \sum_{k \in \text{K-nearest}} X_{kj}
\]

\end{itemize}

\subsubsection*{Multiple Imputation by Chained Equations (MICE)}

Multivariate Imputation by Chained Equations (MICE) is a method used to impute missing values in a dataset by iteratively modeling each variable with missing data as a function of other variables in the dataset. Using LightGBM (Light Gradient Boosting Machine) for MICE involves employing gradient boosting, a powerful machine learning technique, to predict missing values iteratively. LightGBM is particularly efficient for this task due to its ability to handle large datasets and its fast training speed. 
\\
\\
\textbf{Mathematical Formulation}:
\begin{itemize}
    \item Let $X = [X_1, X_2, \dots, X_d] \in \mathbb{R}^{n \times d}$ be the dataset with some missing values.

\item MICE models each feature with missing values as a function of the other features. The imputation proceeds in a round-robin fashion, iteratively filling in missing values:
\begin{itemize}
    \item For feature $X_j$ with missing entries, it fits a regression model $f_j$ on $X_{-j}$ (the other features):

\[
X_j^{(t+1)} = f_j(X_{-j} + \epsilon
\] 
where $\epsilon$ is the random error term drawn from the residuals to capture uncertainty.
\end{itemize}

\item This process is iterated for each feature with missing data:
\begin{itemize}
    \item At iteration $t$, for each $j$, estimate $f_j$ and update the imputed values for $X_j$.
    \item Continue for $T$ iterations to generate multiple imputed datasets $X^{(1)}, X^{(2)}, \dots, X^{(T)}$.
\end{itemize}
\item Finally, the results across the multiple imputed datasets are pooled to account for the uncertainty in the imputations. The final estimate for a parameter $\theta$ is obtained by combining estimates across the imputed datasets:

\[
\hat{\theta} = \frac{1}{T} \sum_{t=1}^{T} \hat{\theta}^{(t)}
\]

\end{itemize}

\subsubsection*{Generative Adversarial Imputation Networks (GAIN)}

Generative Adversarial Imputation Networks (GAIN) is a technique used for imputing missing values in a dataset using deep learning and generative adversarial networks (GAN). GAIN is particularly effective for imputing missing values in high-dimensional datasets with complex dependencies between variables.
\\
\\
\textbf{Mathematical Formulation}:
\begin{itemize}
    \item Let $X \in \mathbb{R}^{n \times d}$ be the dataset with missing values, and $M \in \{0, 1\}^{n \times d}$ be a binary mask, where $M_{ij} = 0$ indicates a missing value and $M_{ij} = 1$ indicates a known value.
    \item GAIN uses two neural networks:
    \begin{itemize}
        \item Generator $G(X, M, Z)$: Takes input data $X$, mask $M$, and random noise $Z$, and generates imputed values for missing entries.
        \item Discriminator $D(X, \tilde{X}, M)$: Takes the original data $X$, the imputed data $\tilde{X}$ (a mix of $X$ and $G(X)$), and the mask $M$, and tries to distinguish between real and imputed values.
    \end{itemize}
\item The generator minimizes the mean squared error (MSE) for the imputed values:

\[
L_G = \mathbb{E} \left[ M \odot (X - G(X, M, Z))^2 \right]
\]

\item 
The discriminator maximizes the probability of correctly identifying real versus imputed data:

\[
L_D = \mathbb{E} \left[ \log D(M \odot X + (1 - M) \odot G(X, M, Z)) \right]
\]

\end{itemize}

\subsection*{ }
\label{oversamplingsection}
\subsection*{C. Oversampling Methods}
\subsubsection*{Random Oversampling Example (ROSE)}
Random Oversampling Examples (ROSE) addresses class imbalance by randomly replicating instances from the minority classes. \\ \\
\textbf{Mathematical Formulation:}

Let $X \in \mathbb{R}^d$ represent the feature vector, and let $Y \in \{0, 1\}$ represent the class label, with class 1 being the minority class. The goal is to generate synthetic data points for the minority class by perturbing the existing minority class instances.
\begin{itemize}
    \item Given a minority class instance $X_i$, generate a synthetic example $\tilde{X}$ by adding a small perturbation:
   \[
   \tilde{X} = X_i + \epsilon
   \]
   where $\epsilon$ is drawn from a kernel density estimate (KDE) of the distribution of $X_i$. Typically, $\epsilon \sim N(0, \Sigma)$ is sampled from a multivariate Gaussian distribution with mean zero and covariance matrix $\Sigma$, which controls the spread of the generated samples.
\item This process is repeated for all instances of the minority class to produce new synthetic points, which are added to the dataset to balance the class distribution.

\end{itemize}

\subsubsection*{Synthetic Minority Oversampling Technique (SMOTE)}
Unlike random oversampling, which simply duplicates instances from the minority class, SMOTE generates synthetic samples for the minority class by interpolating between existing minority class instances. By creating synthetic samples instead of simply duplicating existing minority class instances, SMOTE helps to increase the diversity of the minority class and mitigate the risk of overfitting. 
\\
\\
\textbf{Mathematical Formulation:}

Let $X_i \in \mathbb{R}^d$ be an instance from the minority class, and $X_{NN(i)}$ be one of its $k$-nearest neighbors, where $NN(i)$ denotes the set of nearest neighbors of $X_i$ in the minority class.
\begin{itemize}
    \item For each minority class instance $X_i$, choose one of its $k$-nearest neighbors $X_{NN(i)}$ at random.
\item  Generate a synthetic sample $\tilde{X}$ by interpolating between $X_i$ and $X_{NN(i)}$ as follows:
   \[
   \tilde{X} = X_i + \lambda (X_{NN(i)} - X_i)
   \]
   where $\lambda \in [0, 1]$ is a random number drawn from a uniform distribution.
\item The newly generated point $\tilde{X}$ lies on the line segment between $X_i$ and its neighbor $X_{NN(i)}$, effectively creating new synthetic data points within the minority class.

\end{itemize}

\subsubsection*{Borderline SMOTE}
Borderline SMOTE is an extension of SMOTE designed to further address class imbalance in datasets. While SMOTE generates synthetic samples uniformly across the minority class, Borderline SMOTE focuses specifically on those instances that are close to the decision boundary — often the most difficult to classify. It primarily generates synthetic examples near these borderline cases, aiming to strengthen the decision boundary for the classifier. \\ \\

\textbf{Mathematical Formulation}

Let $X_i \in \mathbb{R}^d$ be a minority class instance, and $X_{NN(i)}$ be one of its $k$-nearest neighbors, where $NN(i)$ denotes the set of nearest neighbors of $X_i$.

\begin{itemize}
    \item For each minority class instance $X_i$, we count the number of its $k$-nearest neighbors that belong to the majority class. Based on this count, each minority class instance is classified into one of three categories:
    \begin{align*}
    \text{Noise:} \quad & n = k \\
    \text{Danger:} \quad & \frac{k}{2} < n < k \\
    \text{Safe:} \quad & 0 \leq n \leq \frac{k}{2}
    \end{align*}
    \item Synthetic samples are generated only for the instances classified as "danger." One of their $k$-nearest neighbors, $X_{NN(i)}$, is randomly selected.
    \item For these "danger" zone instances, a synthetic sample $\tilde{X}$ is created by interpolation:
    \[
    \tilde{X} = X_i + \lambda (X_{NN(i)} - X_i)
    \]
    where $\lambda \sim U(0, 1)$ is a random variable uniformly drawn between 0 and 1. This linear interpolation creates a new synthetic data point between $X_i$ and its neighbor, thus strengthening the decision boundary.
\end{itemize}

\textbf{Mathematical Formulation}

Let $X_i \in \mathbb{R}^d$ be a minority class instance, and $X_{NN(i)}$ be one of its $k$-nearest neighbors, where $NN(i)$ denotes the set of nearest neighbors of $X_i$.
\begin{itemize}
    \item For each minority class instance $X_i$, randomly select one of its $k$-nearest neighbors $X_{NN(i)}$.
\item Generate a synthetic sample $\tilde{X}$ by interpolation:
   \[
   \tilde{X} = X_i + \lambda (X_{NN(i)} - X_i)
   \]
   where $\lambda \sim U(0, 1)$ is a random number uniformly drawn between 0 and 1. This linear interpolation creates a new synthetic data point between $X_i$ and its neighbor.

\end{itemize}

\subsubsection*{Adaptive Synthetic Sampling (ADASYN)}

ADASYN is another extension of SMOTE. It focuses on generating synthetic data points adaptively, creating more samples in regions where the minority class is harder to classify. ADASYN assigns a weight to each minority instance based on the density of majority class neighbors, generating more synthetic points for instances surrounded by more majority class examples. This ensures that areas with higher class imbalance receive more synthetic data, helping to improve classifier performance in those challenging regions. \\ \\

\textbf{Mathematical Formulation}

Let $X_i \in \mathbb{R}^d$ represent a minority class instance, and $NN(i)$ be the set of $k$-nearest neighbors of $X_i$. Define $r_i$ as the ratio of majority class instances among the $k$-nearest neighbors of $X_i$:

\[
r_i = \frac{\text{Number of majority class neighbors of } X_i}{k}
\]
\begin{itemize}
    \item Normalize the difficulty ratio $r_i$ to obtain a probability distribution $p_i$:
   \[
   p_i = \frac{r_i}{\sum_{i=1}^{n} r_i}
   \]
   where $n$ is the number of minority class instances.
\item For each minority class instance $X_i$, generate $G_i = p_i \cdot G$ synthetic examples, where $G$ is the total number of synthetic examples to be generated.
\item For each $X_i$, generate a synthetic sample $\tilde{X}$ as in SMOTE by selecting a random neighbor $X_{NN(i)}$ and interpolating:
   \[
   \tilde{X} = X_i + \lambda (X_{NN(i)} - X_i)
   \]
   where $\lambda \sim U(0, 1)$.

\end{itemize}

\subsection*{D. Machine Learning Models}
\label{ml}

\subsubsection*{Logistic Regression}

Logistic Regression is a statistical method used for binary classification that models the probability of a binary outcome based on one or more predictor variables. It uses the logistic function to establish the relationship between the dependent variable \( Y \) and the independent variables \( \mathbf{X} = (X_1, X_2, \ldots, X_p) \). The model is defined as follows:

Let \( Y \) be the binary outcome variable, where \( Y = 1 \) indicates the presence of a condition (e.g., success) and \( Y = 0 \) indicates its absence (e.g., failure). The probability of \( Y \) being 1 given \( \mathbf{X} \) is expressed as:

\[
P(Y = 1 | \mathbf{X}) = \sigma(\mathbf{X}^T \mathbf{\beta}) = \frac{1}{1 + e^{-\mathbf{X}^T \mathbf{\beta}}}
\]

where:
\begin{itemize}
    \item \( \sigma(z) \) is the logistic (sigmoid) function,
    \item \( \mathbf{\beta} = (\beta_0, \beta_1, \ldots, \beta_p) \) is a vector of coefficients (parameters) to be estimated,
    \item \( \mathbf{X}^T \) is the transpose of the vector of independent variables.
\end{itemize}

The odds of the event occurring can be defined as:

\[
\text{Odds}(Y = 1 | \mathbf{X}) = \frac{P(Y = 1 | \mathbf{X})}{P(Y = 0 | \mathbf{X})} = \frac{\sigma(\mathbf{X}^T \boldsymbol{\beta})}{1 - \sigma(\mathbf{X}^T \boldsymbol{\beta})}
\]

To estimate the parameters \( \mathbf{\beta} \), the method of maximum likelihood estimation (MLE) is typically used, which seeks to minimize binary cross-entropy loss defined as:

\[
L(\mathbf{\beta}) = -\frac{1}{n} \sum_{i=1}^{n} \left[ Y_i \log(\hat{Y}_i) + (1 - Y_i) \log(1 - \hat{Y}_i) \right]
\]

where:
\begin{itemize}
    \item \( Y_i \) is the actual binary outcome for the \( i \)-th observation,
    \item \( \hat{Y}_i = P(Y_i = 1 | \mathbf{X}_i) = \sigma(\mathbf{X}_i^T \mathbf{\beta}) \) is the predicted probability that \( Y_i = 1 \).
\end{itemize}

The terms \( \log(\hat{Y}_i) \) and \( \log(1 - \hat{Y}_i) \) penalize the model when the predicted probabilities deviate from the actual outcomes, ensuring that the logistic regression model learns to provide accurate probability estimates for the binary class.

\subsubsection*{Random Forest}
Random Forest is an ensemble learning method primarily used for classification and regression tasks. It builds multiple decision trees during training and outputs the mode of the classes (for classification) or the mean prediction (for regression) of the individual trees. This method is based on the principle of aggregating the predictions of several models to improve overall performance and reduce overfitting.

\subsubsubsection*{Basic Concepts}

Let:
\[
\mathcal{D} = \{ (x_1, y_1), (x_2, y_2), \ldots, (x_n, y_n) \}
\]
be a training dataset, where \( x_i \in \mathbb{R}^d \) are the input features and \( y_i \) are the corresponding target values (labels). Let \( N \) be the number of trees in the Random Forest.

\subsubsubsection*{Bootstrap Aggregating (Bagging)}

Random Forest employs a technique called Bootstrap Aggregating, or \textbf{Bagging}, to create diverse trees. For each tree \( t \) (where \( t = 1, 2, \ldots, N \)):
\begin{enumerate}
    \item \textbf{Sampling}: Generate a bootstrap sample \( \mathcal{D}_t \) by randomly sampling with replacement from \( \mathcal{D} \). This sample typically has the same size as \( \mathcal{D} \) but may contain duplicate instances.
    \[
    \mathcal{D}_t = \{ (x_{i_1}, y_{i_1}), (x_{i_2}, y_{i_2}), \ldots, (x_{i_m}, y_{i_m}) \}
    \]
    where \( m \) is the size of the sample.
    
    \item \textbf{Training}: Train a decision tree \( f_t(x) \) on the bootstrap sample \( \mathcal{D}_t \).
\end{enumerate}

\subsubsubsection*{Tree Construction}

During the construction of each tree:
\begin{itemize}
    \item \textbf{Feature Selection}: For each split in the decision tree, a random subset of features \( \mathcal{F} \subset \{1, 2, \ldots, d\} \) is chosen. Let \( m \) be the number of features selected (commonly \( m = \sqrt{d} \)).
    
    \item \textbf{Best Split}: The best split is determined by evaluating all possible splits among the selected features. The impurity measure can be based on criteria such as Gini impurity or entropy for classification, or mean squared error for regression.
    
    The Gini impurity for a split can be defined as:
    \[
    Gini(p) = 1 - \sum_{j=1}^{C} p_j^2
    \]
    where \( p_j \) is the proportion of class \( j \) in the dataset, and \( C \) is the total number of classes.
\end{itemize}

\subsubsubsection*{Aggregating Predictions}

Once all \( N \) trees have been trained, predictions for a new instance \( x \) are made by aggregating the predictions from each tree:
\begin{itemize}
    \item For classification tasks:
    \[
    \hat{y} = \text{mode}(f_1(x), f_2(x), \ldots, f_N(x))
    \]
    
    \item For regression tasks:
    \[
    \hat{y} = \frac{1}{N} \sum_{t=1}^{N} f_t(x)
    \]
\end{itemize}

\subsubsubsection*{Loss Function}

During training, the Random Forest minimizes a loss function to ensure the quality of the predictions. For classification problems, the most commonly used loss function is the cross-entropy loss, defined as:

\[
L(y, \hat{y}) = - \frac{1}{n} \sum_{i=1}^{n} \sum_{c=1}^{C} y_{i,c} \log\hat{y}_{i,c}
\]

where:
\begin{itemize}
    \item \( y \) is the true label (one-hot encoded) for each observation,
    \item \( \hat{y} \) is the predicted probability for each class \( c \),
    \item \( n \) is the number of observations, and
    \item \( C \) is the number of classes.
\end{itemize}

For regression problems, the mean squared error (MSE) is often used as the loss function:

\[
L(y, \hat{y}) = \frac{1}{n} \sum_{i=1}^{n} (y_i - \hat{y}_i)^2
\]

where:
\begin{itemize}
    \item \( y_i \) is the actual target value for the \( i \)-th observation,
    \item \( \hat{y}_i \) is the predicted value from the Random Forest.
\end{itemize}

\subsubsection*{XGBoost}
XGBoost is a powerful ensemble method that combines decision trees and gradient boosting. By minimizing a regularized loss function and efficiently fitting trees to residuals, XGBoost delivers high predictive performance and robust generalization.

\subsubsubsection*{Basic Concepts}

Let:
\[
\mathcal{D} = \{(x_i, y_i)\}_{i=1}^n
\]
be the same training dataset defined above, where \( x_i \in \mathbb{R}^d \) are the input features and \( y_i \) are the corresponding target values (labels). Let \( T \) be the total number of trees in the ensemble.

\subsubsubsection*{Loss Function}

The objective of XGBoost is to minimize the following regularized loss function:
\[
\mathcal{L} = \sum_{i=1}^{n} L(y_i, \hat{y}_i) + \sum_{j=1}^{T} \Omega(f_j)
\]
where:
\begin{itemize}
    \item \( L(y_i, \hat{y}_i) \) is the loss function that measures the difference between the true label \( y_i \) and the predicted label \( \hat{y}_i \).
    \item \( \Omega(f_j) \) is the regularization term for the \( j \)-th tree, which helps control the complexity of the model and prevent overfitting.
\end{itemize}

The regularization term is typically defined as:
\[
\Omega(f_j) = \gamma T + \frac{1}{2} \lambda ||w_j||^2
\]
where:
\begin{itemize}
    \item \( T \) is the number of leaves in the tree.
    \item \( w_j \) are the leaf weights.
    \item \( \gamma \) and \( \lambda \) are hyperparameters controlling the complexity of the tree.
\end{itemize}

\subsubsubsection*{Gradient Boosting Framework}

The boosting process involves iteratively adding trees to minimize the loss function. Let \( \hat{y}_i^{(t)} \) be the prediction after \( t \) trees. The \( (t+1) \)-th tree is fitted to the residual errors of the predictions from the previous \( t \) trees:
\[
r_i^{(t)} = -\frac{\partial L(y_i, \hat{y}_i^{(t)})}{\partial \hat{y}_i^{(t)}}
\]
This equation computes the gradient of the loss function with respect to the predictions, providing the pseudo-residuals \( r_i^{(t)} \) that the new tree will aim to predict.

\subsubsubsection*{Tree Structure and Splitting}

XGBoost uses a tree structure where each node is split based on the input features to minimize the loss function. The split is chosen to maximize the gain, defined as:
\[
Gain = \frac{1}{2} \left( \frac{(G_L)^2}{H_L + \lambda} + \frac{(G_R)^2}{H_R + \lambda} - \frac{(G)^2}{H + \lambda} \right) - \gamma
\]
where:
\begin{itemize}
    \item \( G \) and \( H \) are the sum of gradients and hessians (second derivatives) of the loss function for the observations in the left and right child nodes, respectively.
    \item \( G_L \) and \( G_R \) are the gradients for the left and right splits.
    \item \( H_L \) and \( H_R \) are the hessians for the left and right splits.
    \item \( \lambda \) is a regularization parameter, and \( \gamma \) is the minimum loss reduction required to make a further partition.
\end{itemize}

\subsubsubsection*{Final Prediction}

The final prediction from XGBoost is the sum of the predictions from all \( T \) trees:
\[
\hat{y}_i = \sum_{t=1}^{T} f_t(x_i)
\]
where \( f_t(x_i) \) is the prediction from the \( t \)-th tree.

\subsubsection*{LightGBM}
LightGBM is a powerful gradient boosting framework that combines efficient training techniques, such as GOSS and histogram-based algorithms, with a leaf-wise growth strategy to enhance predictive performance. Its ability to handle large datasets and high-dimensional features makes it a popular choice for a wide range of machine learning applications.

\subsubsubsection*{Loss Function}

The objective of LightGBM is to minimize the following regularized loss function:
\[
\mathcal{L} = \sum_{i=1}^{n} L(y_i, \hat{y}_i) + \sum_{j=1}^{T} \Omega(f_j)
\]
where:
\begin{itemize}
    \item \( L(y_i, \hat{y}_i) \) is the loss function that measures the discrepancy between the true label \( y_i \) and the predicted label \( \hat{y}_i \).
    \item \( \Omega(f_j) \) is the regularization term for the \( j \)-th tree to prevent overfitting and control complexity.
\end{itemize}

The regularization term can be defined as:
\[
\Omega(f_j) = \gamma T + \frac{1}{2} \lambda ||w_j||^2
\]
where:
\begin{itemize}
    \item \( T \) is the number of leaves in the tree.
    \item \( w_j \) are the leaf weights.
    \item \( \gamma \) and \( \lambda \) are hyperparameters that control the complexity of the model.
\end{itemize}

\subsubsubsection*{Gradient-Based One-Side Sampling (GOSS)}

One of the unique features of LightGBM is the Gradient-Based One-Side Sampling (GOSS) technique, which selects data instances based on their gradients. First, the instances with larger gradients (higher errors) are kept for training. Second, random subset of instances with smaller gradients is also selected to ensure that the training process is efficient while maintaining accuracy.

This method reduces the number of data instances used for constructing the tree while preserving the most informative samples.

\subsubsubsection*{Histogram-Based Algorithm}

LightGBM uses a histogram-based algorithm to speed up the process of finding optimal split points for continuous features. First, continuous feature values are bucketed into discrete bins. Then, the algorithm computes the best split points based on these bins instead of the raw feature values, significantly reducing computation time.

The gain from a split is computed as:
\[
Gain = \frac{1}{2} \left( \frac{(G_L)^2}{H_L + \lambda} + \frac{(G_R)^2}{H_R + \lambda} - \frac{(G)^2}{H + \lambda} \right) - \gamma
\]
where:
\begin{itemize}
    \item \( G \) and \( H \) are the gradients and hessians (second derivatives) of the loss function for the observations in the split.
    \item \( G_L \) and \( G_R \) are the gradients for the left and right splits.
    \item \( H_L \) and \( H_R \) are the hessians for the left and right splits.
\end{itemize}

\subsubsubsection*{Leaf-wise Growth Strategy}

Unlike traditional boosting methods that grow trees level-wise, LightGBM employs a leaf-wise growth strategy:
- The algorithm grows the tree by adding leaves with the highest potential gain, which often leads to deeper trees but improves model accuracy.

The new leaf \( l \) is added based on the highest gain calculated for possible splits.

\subsubsubsection*{Final Prediction}

The final prediction from LightGBM is obtained by summing the predictions from all \( T \) trees:
\[
\hat{y}_i = \sum_{t=1}^{T} f_t(x_i)
\]
where \( f_t(x_i) \) is the prediction from the \( t \)-th tree.

\subsubsection*{CatBoost}
CatBoost is a powerful gradient boosting framework that excels at handling categorical features and offers robust performance through ordered boosting. Its unique approach to feature encoding and efficient training methods make it a preferred choice for many machine learning applications, particularly when dealing with datasets containing categorical variables.

\subsubsubsection*{Loss Function}

The objective of CatBoost is to minimize the following regularized loss function:
\[
\mathcal{L} = \sum_{i=1}^{n} L(y_i, \hat{y}_i) + \sum_{j=1}^{T} \Omega(f_j)
\]
where:
\begin{itemize}
    \item \( L(y_i, \hat{y}_i) \) is the loss function measuring the discrepancy between the true label \( y_i \) and the predicted label \( \hat{y}_i \).
    \item \( \Omega(f_j) \) is the regularization term for the \( j \)-th tree to prevent overfitting and control complexity.
\end{itemize}

The regularization term is typically defined as:
\[
\Omega(f_j) = \gamma T + \frac{1}{2} \lambda ||w_j||^2
\]
where:
\begin{itemize}
    \item \( T \) is the number of leaves in the tree.
    \item \( w_j \) are the leaf weights.
    \item \( \gamma \) and \( \lambda \) are hyperparameters controlling the complexity of the model.
\end{itemize}

\subsubsubsection*{Handling Categorical Features}

One of the distinguishing features of CatBoost is its ability to handle categorical variables without the need for extensive preprocessing, such as one-hot encoding. CatBoost employs a technique called ordered boosting, which uses the information from the categorical features while avoiding target leakage.

To encode categorical features, CatBoost computes the average target value for each category in a way that preserves the order of the observations. For a categorical feature \( C \):
\[
C_{new} = \frac{\sum_{i \in \text{cat}} y_i + \alpha}{n_{\text{cat}} + \beta}
\]
where:
\begin{itemize}
    \item \( n_{\text{cat}} \) is the count of observations in the category.
    \item \( y_i \) are the target values for observations in the category.
    \item \( \alpha \) and \( \beta \) are smoothing parameters to avoid overfitting.
\end{itemize}

\subsubsubsection*{Gradient Boosting Framework}

The boosting process involves iteratively adding trees to minimize the loss function. Let \( \hat{y}_i^{(t)} \) be the prediction after \( t \) trees. The \( (t+1) \)-th tree is fitted to the residual errors of the predictions from the previous \( t \) trees:
\[
r_i^{(t)} = -\frac{\partial L(y_i, \hat{y}_i^{(t)}}{\partial \hat{y}_i^{(t)}}
\]
This equation computes the gradient of the loss function with respect to the predictions, providing the pseudo-residuals \( r_i^{(t)} \) that the new tree will aim to predict.

\subsubsubsection*{Tree Construction and Splitting}

During the construction of each tree, CatBoost selects the best split for each node based on the gradients and hessians:
\[
Gain = \frac{1}{2} \left( \frac{(G_L)^2}{H_L + \lambda} + \frac{(G_R)^2}{H_R + \lambda} - \frac{(G)^2}{H + \lambda} \right) - \gamma
\]
where:
\begin{itemize}
    \item \( G \) and \( H \) are the gradients and hessians (second derivatives) of the loss function for the observations in the split.
    \item \( G_L \) and \( G_R \) are the gradients for the left and right splits.
    \item \( H_L \) and \( H_R \) are the hessians for the left and right splits.
\end{itemize}

\subsubsubsection*{Final Prediction}

The final prediction from CatBoost is obtained by summing the predictions from all \( T \) trees:
\[
\hat{y}_i = \sum_{t=1}^{T} f_t(x_i)
\]
where \( f_t(x_i) \) is the prediction from the \( t \)-th tree.

\end{document}